\documentclass[epj,referee]{svjour}
\usepackage{graphics}
\begin{document}
\title{A Deterministic Approach to the Synchronization of Cellular Automata.}
\subtitle{}
\author{Jos\'e Garc\'ia \inst{1}\inst{2}  \and Pedro Garc\'ia\inst{2}
}                     
%
%
\institute{Centro de F\'{\i}sica Te\'orica y Computacional, 
Escuela de F\'{\i}sica,
Facultad de Ciencias, Universidad Central de Venezuela,
Caracas, Venezuela. \and Laboratorio de Sistemas Complejos, 
Departamento de F\'{\i}sica Aplicada,
Facultad de Ingenier\'{\i}a, Universidad Central de Venezuela, Caracas, Venezuela.}
\date{Received: date / Revised version: date}
%
\abstract{
In this work we introduce a deterministic scheme of synchronization of linear and nonlinear cellular automata (CA) with complex behavior, connected through a master-slave coupling. By using a definition of Boolean derivative, we use the linear approximation of the automata to determine a function of coupling that promotes synchronization without perturbing all the sites of the slave system.
} 
\maketitle
\section{Introduction}
\label{intro}
The spontaneous and synchronous evolution of extended systems or parts of them is an increasingly frecuent 
phenomenon in disciplines ranging from biology to meteorology\cite{Buck,Mirollo,Pikovsky,Strogatz,Kaka,Mancoff,Yair,Adioui}. In several of these cases, the systems are made up of finite, discrete and interrelated  parts.
Nevertheless, the major models used to study the synchronization of extended systems involve continuous systems represented by Partial Differential Equations\cite{Kocarev,Bragard,Boccaletti,Amengual,Garcia-Acosta-Leiva} or partially discrete models given by Coupled Maps Networks, see for instance \cite{Zanette-Morelli} and references therein. In both cases the problem of synchronization of identical and non-identical chaotic systems has been addressed.\\
There is a remarkably small number are cases where the model is represented by a totally discrete system such as CAs\cite{Morelli-Zanette,Bagnoli-Rechtman,Sanchez-Lopez-Ruiz,Rouquier,Urias-Salazar-Ugalde,Salazar-Ugalde-Urias,Rouquier,Bagnoli-ElYacoubi-Rechtman}. The before mentioned references are, as far as we know, the only cases. Particularly, in \cite{Morelli-Zanette} the authors study the synchronization between two stochastically coupled CA with complex behavior, where the coupling mechanism is characterized by a probability $p$. In \cite{Bagnoli-Rechtman}, the authors study the synchronization between randomly coupled totalistic (linear) CA; as in the previous reference the coupling is characterized by a probability $p$ and the synchronization threshold  $p_c$ is related to the maximum Lyapunov exponent of the automata, defined in analogy with continuous dynamical systems. Along the same line is the work of Sanchez et. al.\cite{Sanchez-Lopez-Ruiz}. In this work the same stochastic coupling scheme is used to search for stable patterns in the evolution of the automata. Rouquier in \cite{Rouquier} carries out an exhaustive experimental study of synchronization in CAs using the coupling function proposed by \cite{Morelli-Zanette}. Finally, in a very recent publication\cite{Bagnoli-ElYacoubi-Rechtman}, F. Bagnoli et. al. use the Jacobian matrix of the CA to estimate a probability that later they use to design a coupling scheme to reach synchronization or achieve control.\\ 
On the other hand, in references \cite{Urias-Salazar-Ugalde,Salazar-Ugalde-Urias} the authors derive conditions for synchronization using a deterministic coupling, but only in the case of totalistic automata.\\
The aim of this work is twofold. Firstly we introduce a deterministic scheme for coupling Class III, linear and nonlinear CAs, coupled by a master-slave scheme. The strategy is based on a sort of linear approximation of the automata by means of a definition of Boolean derivative\cite{Vichniac}. And secondly, we propose criteria to minimize the intensity of the coupling, in the sense of the number of coupling sites.\\
The article is organized as follows: in section 2 we introduce briefly the CA, the notion of the Boolean derivative, set the problem of synchronization between CAs and establish conditions for synchronization in unidirectionally coupled  CAs. In section 3, we show the performance of the strategy using examples of Class III nonlinear CAs, and finally in section 4 we make some concluding remarks.

\section{Linear approximation of nonlinear cellular automata and synchronization}
Cellular automatas are dynamical systems where space, time and state variables belong to discrete sets. More formally, a CA is a $4$-tuple $( {\cal L, S, N}, f )$, where $\cal L$ is a $d$-dimensional lattice of cells, $\cal S$ denotes the set of the $k$ accessible estates of an individual cell, $\cal N$ is a mapping defining the neighborhood of each cell and $f$ is a local transition rule, that gives the state of the $i$-th cell in the time $n+1$ given state of it neighborhood in the time $n$, it is
\begin{equation}
s_{n+1}^i=f(N(i))
\end{equation}

Despite it simplicity, CAs exhibit an unsuspected ability to emulate, from qualitative\cite{Greenberg-Hastings} and quantitative\cite{Weimar-Boon} points of view, complex extended systems and offer an alternative to model a great variety of systems, such as: fluid dynamics, forest fires, ecosystems or epidemics. 

Here, we shall restrict ourselves to the case of one-dimensional binary CAs of $N$ cells with neighborhood of size ($r$) equal to 3, defined as {\it elementary} by Wolfram in \cite{Wolfram}. In these models,  $\cal L$ is a linear array where each cell adopts one of two possible states in ${\cal S}=\{ 0,1 \}$ and the dynamical rule is the form $s_{n+1}^i=f(s_{n}^{i-1},s_{n}^i,s_{n}^{i+1})$, it means that, ${\cal N}(i)\to\{ s_{n}^{i-1},s_{n}^i,s_{n}^{i+1} \}$. In this case, there are $k^{k^r}=256$ possible rules $f$. If $\{ b_i \}$  is a binary string, corresponding to the result of applying $f$ to each different set of 3-tuples  $(s^{i-1},s^i,s^{i+1})$ sorted in descending order, it is possible to use the Wolfram$^{\prime}$s naming scheme\cite{Wolfram} to assign, to each rule, a label given by the decimal representation of the binary sequence $\{ b_i \}$.  

Rules of elementary automatas were classified qualitatively by Wolfram\cite{Wolfram} according to its asymptotic behavior as Class I to  IV, if the automata evolution asymptotically goes to: a constant state, an isolated periodic segments, chaotic regime or isolated chaotic segments, respectively.

In contrast with the continuous case, the elements of the Class III can be linear or nonlinear CA. The linear are the rules that can be written as
\begin{equation}
f(s^{i-1}_n, s^{i}_n,s^{i+1}_n) = \sum_{j=i-1}^{i+1} a_j s^j_{n} ~~mod(2) ~~~~~~a_j \in \{ 0,1 \}.
\end{equation}
\noindent
In the case of elementary CA this class contain 16 elements and the rest are nonlinear.   

For an array of size $N$ it is possible to define the automata state as ${\bf S}_n= \{ s^1_n,s^2_n, \dots, s^N_n\}$ whose evolution is governed by 
\begin{equation}
{\bf S}_{n+1} = F({\bf S_n})
\label{global-rule}
\end{equation}
\noindent
where $F$ is a global rule induced by the local rule $f$.

In our problem, we deal with two identical Class III automatas ${\bf S}_{n+1} = F({\bf S}_n)$ (the driver system) and $\tilde {\bf S}_{n+1} = F(\tilde  {\bf S}_n)$ (the driven system) with periodic boundary conditions and different initial conditions  ${\bf S}_0$ and  $\tilde  {\bf S}_0$.

In analogy with an usual strategy in the continuous case, we connect the systems by mean of an unidirectional linear coupling that penalize the separation between the estates of the driver and driven systems in the form
\begin{eqnarray}
{\bf S}_{n+1} & = & F({\bf S_n}) \nonumber \\
\tilde {\bf S}_{n+1} & = & F(\tilde  {\bf S}_n) \oplus {\bf K} \odot ({\bf S}_n \oplus \tilde  {\bf S}_n)
\end{eqnarray}
\noindent
where $\oplus$ and $\odot$ denotes the binary operations OR exclusive and AND respectively and  ${\bf K}$ is a coupling matrix that define the coupling, this is, the number and location where the master system perturb to the slave system.

The evolution of the difference automata defined as $\Delta_{n+1}={\bf S}_{n+1} \oplus \tilde {\bf S}_{n+1}$, is given by:
\begin{eqnarray}
\Delta_{n+1} & = & F({\bf S_n}) \oplus F(\tilde  {\bf S}_n) \oplus {\bf K} \odot ({\bf S}_n \oplus \tilde  {\bf S}_n) \nonumber \\
 & \approx & F({\bf S_n}) \oplus F({\bf S}_n) \oplus J({\bf S}_n) \odot \Delta_n \oplus {\bf K} \odot \Delta_{n} \nonumber \\
 & \approx &  \left( J({\bf S}_n) \oplus {\bf K} \right) \odot \Delta_n
\label{sync}
\end{eqnarray}
\noindent
where $F(\tilde  {\bf S}_n)$ is linearly approximated as in \cite{Bagnoli-Rechtman} and the Jacobian matrix of $F$, $J({\bf S}_n)$ was defined in \cite{Vichniac}, as
\begin{equation}
\frac{\partial F}{\partial s^j} = F(s^i, \dots, s^j, \dots) \oplus F(s^i, \dots, s^j \oplus 1, \dots)
\end{equation}
\noindent
with matrix elements:
\begin{equation}
J_{i,j} = \frac{\partial F_i}{\partial s^j} 
\label{Jacobiano-booleano}
\end{equation}
   
It is worth noting that this Jacobian is a tri-diagonal matrix except for two off-diagonal elements in the corners for periodic boundary condition. This structure could be colloquially explained as follows: each element in the principal diagonal represent a cell, an 1 in this diagonal indicate that this site is affected in the future by a change of it in the present; ones in the column indicate that this site is affected in the future by actual changes in its neighbors and ones in the file indicate that changes in this site affect, in the future, its neighbors.
 
From (\ref{sync}), it is clear that if ${\bf K} = J({\bf S}_n)$ the difference automata converge to zero and the master-slave system synchronizes. Nevertheless, this coupling represents a connection between all elements of the master and slave systems by mean of a coupling function that varies in time. 

In order to design a more efficient coupling function (in the sense of the complexity and intensity of this function) we suppress the time variation approximating $\bf J({\bf S}_n)$ by $\bf \bar J({\bf S}_n,q)$, as:
\begin{equation}
{\bar {\bf J}}({\bf S}_n,q) = H \left( q-\frac{1}{T} \sum_{n=1}^{T} J({\bf S}_n) \right) 
\label{Jacobiano-booleano}
\end{equation}
\noindent
where $H$ is the Heaviside function and $0 \leq q \leq 1$ is a threshold parameter that represents a kind of tolerance in the approximation of the local Jacobian by the average Jacobian. In the particular case of linear automata, $\bar {\bf J}$ coincides with the Exact Jacobian independently of the value of $q$. 

In this way, the resultant Jacobian could  yet have a structure so that its application in (\ref{sync}) results in the  perturbation of all sites in the slave system. In order to minimize the number of coupling sites, we will chose $\bf K$, selecting the coupling sites as the sites where the perturbations are more easily spread, and at the same time, are sites more sensitive to be perturbed by changes in its neighborhood. These are sites ($c$) so that the Jacobian matrix  has 2 or 3 ones in the corresponding row ($J_{c,i}$) or column ($J_{i,c}$).
\begin{equation}
{\bf K} = \left\{
            \begin{array}{lc}
		  {\bar J}_{i,i}(q)     & {\rm si}~~  h \geq 4 \\
                  0                     & {\rm si}~~  h < 4 
            \end{array}
          \right. 
\label{coupling}
\end{equation}
\noindent
where $h$ is
\begin{equation}
h = \left( \sum_{n=-1}^{1} \bar J_{i+n,i}(q) \right) \left( \sum_{n=-1}^{1} \bar J_{i,i+n}(q) \right)
\end{equation}

If $F$ is nonlinear ${\bf S}_{n+1} \neq {\bf J}(q) {\bf S}_{n}$ and of course ${\bf S}_{n+1} \neq \bar {\bf J}(q) {\bf S}_{n}$, nevertheless as we will see, a reiterative application of the perturbation over the slave system can drive its state close to the state of the master system using a suitable value of $q$.
\label{sec:1}

\section{Numerical Results}
\begin{figure}[h]
\resizebox{0.75\columnwidth}{!}{%
\includegraphics{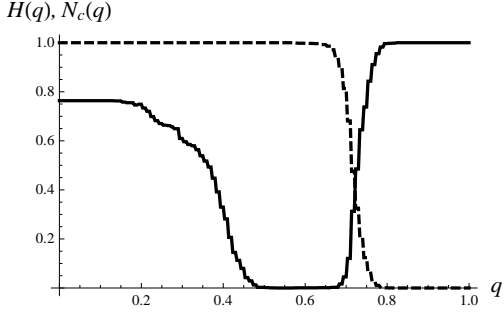}
}
\caption{Normalized Hamming distance $H(q)$ (continuous line) between asymptotic states of the master and the slave systems and Normalized number of coupling sites $N_c(q)$ (dashed line) as a function of $q$, averaged for the case of the automatas 18, 110,126, 146.}
\label{fig1}       
\end{figure}

\begin{figure}[h]
\resizebox{0.75\columnwidth}{!}{%
\includegraphics{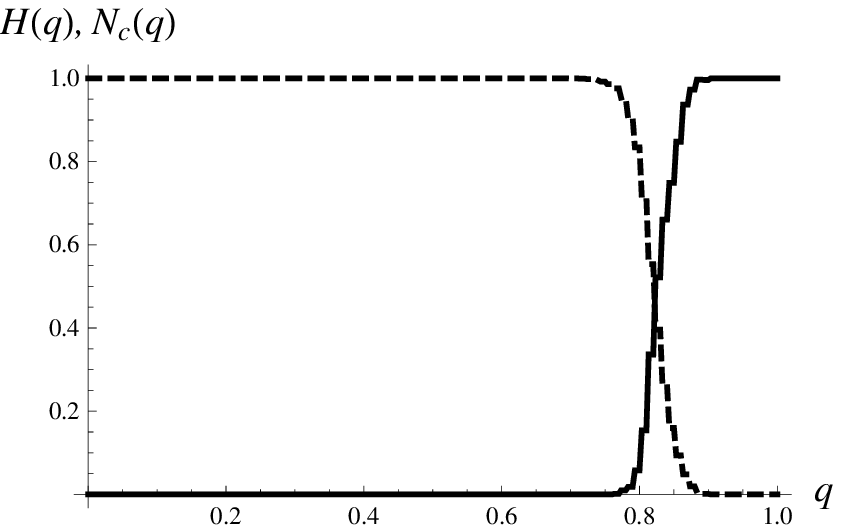}
}
\caption{Normalized Hamming distance $H(q)$ (continuous line) between asymptotic states of the master and the slave systems and Normalized number of coupling sites $N_c(q)$ (dashed line) as a function of $q$, averaged for the case of the automatas 22, 30, 41, 122.}
\label{fig2}
\end{figure}

\begin{figure}[h]
\resizebox{0.75\columnwidth}{!}{%
\includegraphics{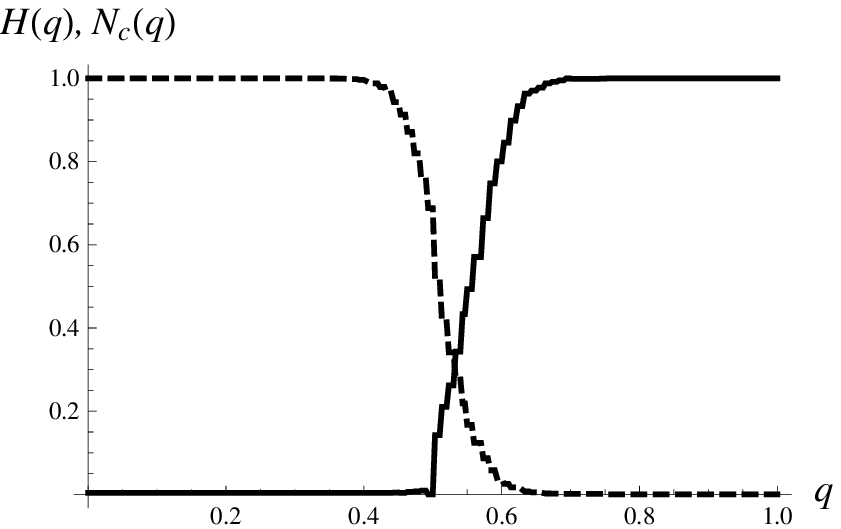}
}
\caption{Normalized Hamming distance $H(q)$ (continuous line) between asymptotic states of the master and the slave systems and Normalized number of coupling sites $N_c(q)$ (dashed line) as a function of $q$, averaged for the case of the automatas 45, 54, 106.}
\label{fig3}
\end{figure}

\begin{figure}[h]
\rotatebox{-90}{
\resizebox{0.75\columnwidth}{!}{%
\includegraphics{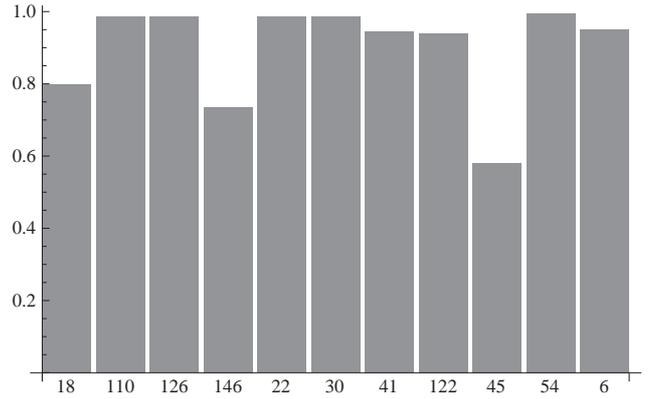}
}}
\caption{Minimum number of sites necessary to reach synchronization for all CAs, averaged over $500$ realizations.}
\label{fig4}
\end{figure}

In order to show the performance of the strategy we carry out experiments with all Wolfram$^{\prime}$s Class III CAs with non-constant Jacobian matrix and present as examples the CAs used in \cite{Bagnoli-Rechtman} {\it i.e.} automatas: $18$, $22$, $30$. $41$, $45$, $54$, $106$, $110$, $122$, $126$ and $146$, with $N=100$. We calculate the normalized Hamming distance $H(q)$ and the fraction of coupling sites, 
\begin{equation}
N_c(q) = \frac{number~of~coupling~sites}{N},
\end{equation}
\noindent
as a function of the parameter $q$, averaged for $500$ randomly chosen initial conditions. Here we can easily identify three types of behavior, that are shown in the Figures (\ref{fig1}), (\ref{fig2}) and (\ref{fig3}): the synchronization occur in a interval of values of $q$ approximately between  $0.5$ and $0.7$, the synchronization occurs in an interval of values of $q$ approximate between y $0.0$ and $0.75$, the synchronization occurs in an interval of values of $q$ approximately between  $0.0$ and $0.5$. In all cases there is a transition from a high number of coupling sites to uncoupling state around the upper limit of the previously mentioned intervals. Figure (\ref{fig4}) shows the minimum number of coupling sites required to achieve synchronization averaged over the 500 randomly chosen initial conditions. As you can see in this figure, there are CAs that should be perturbed in almost all sites in order to reach a synchronized state. This is because the matrix $\bf K$, which determines the number of coupling sites, is the result of averaging $500$ realizations started from randomly chosen states and it is quite likely to depend on the difference between the initial conditions, as suggested in\cite{Garcia-Acosta-Leiva}.

\begin{figure}[h]
\resizebox{0.75\columnwidth}{!}{%
\includegraphics{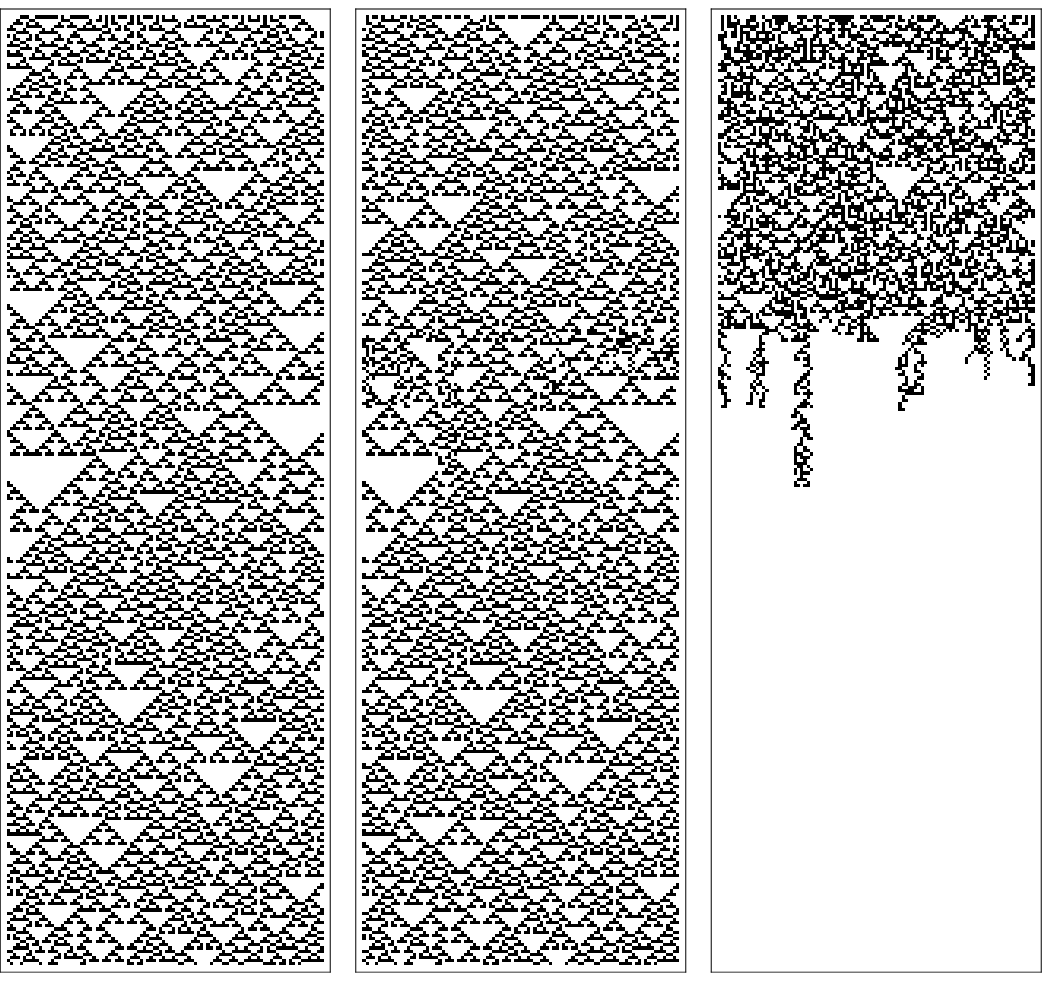}
}
\caption{Evolution of master, slave and difference automata for the rule 22 and $q=0.802$. The horizontal axis represent the space and the time goes from top to bottom}
\label{master-slave22}
\end{figure}

As an specific example, we use the automata rule 22:
\begin{equation}
f(s^{i-1}_n, s^{i}_n,s^{i+1}_n) =  s^{i-1}_n \oplus s^{i}_n \oplus s^{i+1}_n \oplus s^{i-1}_n \odot s^{i}_n \odot s^{i+1}_n 
\end{equation}
\noindent
to show in the Figure (\ref{master-slave22}), the evolution of the master-slave system  when the parameter $q = 0.802$. In this particular case, the number of coupling sites was $N_c = 0.66$ 

Finally, it is important to note that during the numerical experiments we observed that when used as coupling a constant matrix randomly generated, the synchronization is achieved in a small number of attemps.

\label{sec:2}
\section{Final Remarks.} 
We present a deterministic master-slave coupling scheme to achieve synchronization in linear and nonlinear CAs. Here the master, in general, need not act on the the full extent of the slave system, in contrast to probabilistic schemes where it is necessary to monitor the entire slave system all the time, as well as the posibility of disturbing all the sites.

The scheme is valid in linear and non-linear cases. Particularly in the case of linear CA, the Jacobian matrix coincides with the global rule $F$ (\ref{global-rule}); hence the synchronization is reached in one iteration using ${\bf K}$ as defined in (\ref{coupling}).

The strategy presented seems to be useful in the design of strategies to control complex extended systems, that can be modeled by cellular automatas, such as epidemics or forest fires, by a efficient pinning.
\label{sec:3}
%
%

%

\begin{thebibliography}{20}
\bibitem{Buck}  Buck J. Sci. Am. {\bf 234}, 74, (1976).

\bibitem{Mirollo}  Mirollo R. E. and Strogatz S. H. Siam J. Appl. Math. {\bf 50}, 1645, (1990).

\bibitem{Winfree} Winfree A. T. Journal of Theoretical Biology. {\bf 16} 15, (1967).

\bibitem{Pikovsky}  Pikovsky A., Rosenblum M. and Kurths J. Synchronization: A Universal Concept in Nonlinear Sciences.
Cambridge University Press, Cambridge, 2001.

\bibitem{Strogatz} Strogatz S. H. Sync: The Emerging Science of Spontaneous Order, Hyperion, New York, 2003.

\bibitem{Kaka}  Kaka S. F., Pufall M. R., Rippard W. H., Silva T. J. and Russek S. E. Nature. {\bf 437} 389, (2005).

\bibitem{Mancoff}  Mancoff F. B., Rizzo N. D., Engel B. N.  and Tehrani S. Nature. {\bf 437}  393, (2005).

\bibitem{Yair} Yair Y., Aviv R., Price C., Asfur M. and Ravid G. Geophysical Research Abstracts. {\bf 9}  03235, (2007).

\bibitem{Adioui}  Adioui M., Treuil J. P. and Arino O. Ecological Modeling. {\bf 167}  19, (2003).

\bibitem{Kocarev}  Kocarev L., Tasev Z. and Parlitz U. Phys. Rev. Lett. {\bf 79}  51, (1997).

\bibitem{Bragard} Bragard J., Boccaletti S. and Mancini H. Physical Review Letters. {\bf 91}, 064103, 2003).

\bibitem{Boccaletti} Boccaletti S., Bragard J., Arecchi F. T. and Mancini H. Phys. Rev. Lett. {\bf 83} (1999) 536.

\bibitem{Amengual} Amengual A., Hernandez-Garc\'{\i}a E., Montangne R. and San Miguel M. Phys. Rev. Lett. {\bf 78} (1997) 536.

\bibitem{Garcia-Acosta-Leiva} P. Garc\'{\i}a, A. Acosta and H. Leiva. Europhys. Lett, 88 60006, (2009).

\bibitem{Zanette-Morelli} D. Zanette and L. G. Morelli. Int. Jour. Bif. Chaos. {\bf 13}, 16 (2003).

\bibitem{Morelli-Zanette} L. G. Morelli and D. Zanette. Phys. Rev. E. {\bf 58}, R8, (1998).

\bibitem{Bagnoli-Rechtman} F. Bagnoli and R. Rechtman. Phys. Rev. E. {\bf 59}, R1307, (1999).

\bibitem{Sanchez-Lopez-Ruiz} J.R. S\'anchez and R. L\'opez-Ruiz. LNCS 3993, 353, (2006).

\bibitem{Rouquier} J-P. Rouquier. Journ\'ees Automates Cellulaires. 250 (2008).

\bibitem{Bagnoli-ElYacoubi-Rechtman} F. Bagnoli, S. El Yacoubi and R. Rechtman.  LNCS 6350, 188, (2010).

\bibitem{Urias-Salazar-Ugalde} J. Ur\'{\i}as, G. Salazar and E. Ugalde. CHAOS {\bf 8}, 814, (1998). 

\bibitem{Salazar-Ugalde-Urias} G. Salazar, E. Ugalde and J. Ur\'{\i}as. Discrete and Continuous Dynamical Systems. {\bf 13}, 491, (2005).  

\bibitem{Vichniac} G. Y. Vichniac. Physica D. {\bf 45}, 63, (1990). 

\bibitem{Greenberg-Hastings} J. M. Greenberg and S. P. Hastings.  SIAM Journal on Applied Mathematics, {\bf 34}, 515–523 (1978). 

\bibitem{Weimar-Boon} J. Weimark and J-P. Boon. Phys. Rev. E, {\bf 49}, 1749, (1994). 

\bibitem{Wolfram} S. Wolfram. Physica D. {\bf 10}, 1 (1984).



\end{thebibliography}
%
%
%

\end{document}